\documentclass[10pt,a4paper,twoside]{article}
\usepackage{epsfig}
\usepackage{baltlat5}
\usepackage{wrapfig}
\begin{document}
\ \
\vspace{-2.5mm}

\setcounter{page}{527}

\titlehead{Baltic Astronomy, vol.\ts 14, 527--533, 2005.}

\titleb{HIGH VELOCITY SPECTROSCOPIC BINARY ORBITS FROM\\ PHOTOELECTRIC
RADIAL VELOCITIES: BD +82 565\,A}

\begin{authorl}
\authorb{A. Bartkevi\v cius}{1,3} and
\authorb{J. Sperauskas}{1,2}
\end{authorl}

\begin{addressl}
\addressb{1}{Institute of Theoretical Physics and Astronomy,
Vilnius University,\\ Go\v{s}tauto 12, Vilnius, LT-01108, Lithuania}

\addressb{2}{Vilnius University Observatory, \v Ciurlionio~29, Vilnius,
 LT-03100, Lithuania}

\addressb{3}{Department of Theoretical Physics, Vilnius Pedagogical
University, Student\c u~39,\\ Vilnius, LT-08106, Lithuania}
\end{addressl}

\submitb{Received 2005 November 21}

\begin{summary} The spectroscopic orbit of a circumpolar high proper
motion
visual binary BD\,+82 565 A component is determined from 57 CORAVEL
radial velocity measurements.  A short period $P$ = 12.69 d and a
moderate eccentricity $e$ = 0.30 are obtained.  The visual system AB has
a projected spatial separation $\sim$\ 830 AU.  The
system's barycenter velocity $V_0$ = \hbox{--86.7 km/s}, the transverse
velocity $V_t$ = 118.7 km/s and the Galactic spatial velocity components
$U$ = --62.6 km/s, $V$ = --84.1 km/s and $W$ = --84.2 km/s give
evidence that it belongs to the thick disk of the Galaxy.  \end{summary}


\resthead{High velocity spectroscopic binary orbits:
 BD +82 565\,A}{J. Sperauskas, A. Bartkevi\v cius}

\sectionb{1}{INTRODUCTION}
\vskip-9pt
In 1988 we initiated a program of photoelectric radial
velocity measurements of Population II single and binary stars
(Bartkevi\v cius \& Sperauskas 1990, 1994, 1999, 2005; Bartkevi\v
cius et al. 1992; Sperauskas \& Bartkevi\v cius 2002).  The
analysis of the results have led to the discovery of some new
radial velocity variables.  With this publication we start to
publish spectroscopic orbits of the newly discovered binaries.

As a high proper motion star ($\mu$ = 0.37\arcsec per year), BD +82 565
was first mentioned in 1916 (communicated by the Astronomer Royal,
MNRAS, 76, 585).  Its proper motion was determined by comparing the
coordinates given in vol.~III of the Greenwich Astrographic Catalogue
with the coordinates from the Circumpolar Stars Catalogue (Carrington
1857).  Luyten included this star in his high proper motion catalogs LTT
(Luyten 1961) and NLTT (Luyten 1979).  The star is also included in the
Lowell Observatory high proper motion survey as G\,259-37, and there a
very good identification chart is given (Giclas et al. 1970).  As a
double common proper motion star, it was discovered by Luyten (1966).
Slightly different data are given in subsequent Luyten's publications
(Luyten 1967, 1968).  The star is also included in the original Luyten
publication of Double Stars with Common Proper Motion (LDS) as LDS 1894
(Luyten 1969).  In NLTT Luyten gives the angular distance $d$ =
12.0\arcsec\ and {\it PA} = 81.0$\degr$ between the A and B components.
Salim \& Gould (2003) in the Revised NLTT Catalogue presents slightly
different values for the AB system:  $d$ = 11.5\arcsec\ and {\it PA} =
87.9$\degr$ for J2000.0.  Our estimates using the CDS ALADIN interactive
sky atlas are:  $d$ = 11.2$\pm$0.2\arcsec, {\it PA} =
79.5$\pm$0.7$\degr$ for J1950.0 and $d$ = 11.4$\pm$0.1\arcsec\, {\it PA}
= 82.9$\pm$0.3$\degr$ for J2000.0.  To our knowledge, no other
measurements of angular distance and position angle between the
components are known.

\sectionb{2}{IDENTIFICATION}
\vskip-9pt
Equatorial coordinates of the A component of the binary
are $\alpha$(2000.0) = 18$^{\rm h}$47$^{\rm m}$02.6389$^{\rm s}$,
$\delta $(2000.0) = +82$\degr$ 43\arcmin\ 30.260\arcsec.  Five
stars are seen in one arcmin field of the first Palomar
Observatory Sky Survey (POSS I) (Figure 1).  In the second-epoch
Palomar survey the third star is blended by the binary due to its
high proper motion (Figure 2). Identification of the stars taken
from the CDS Simbad is presented in Table 1. Stars 3, 4 and 5 are
optical components.

\vskip3mm

\hbox{
\vtop{\hsize61mm
\centerline{\psfig{figure=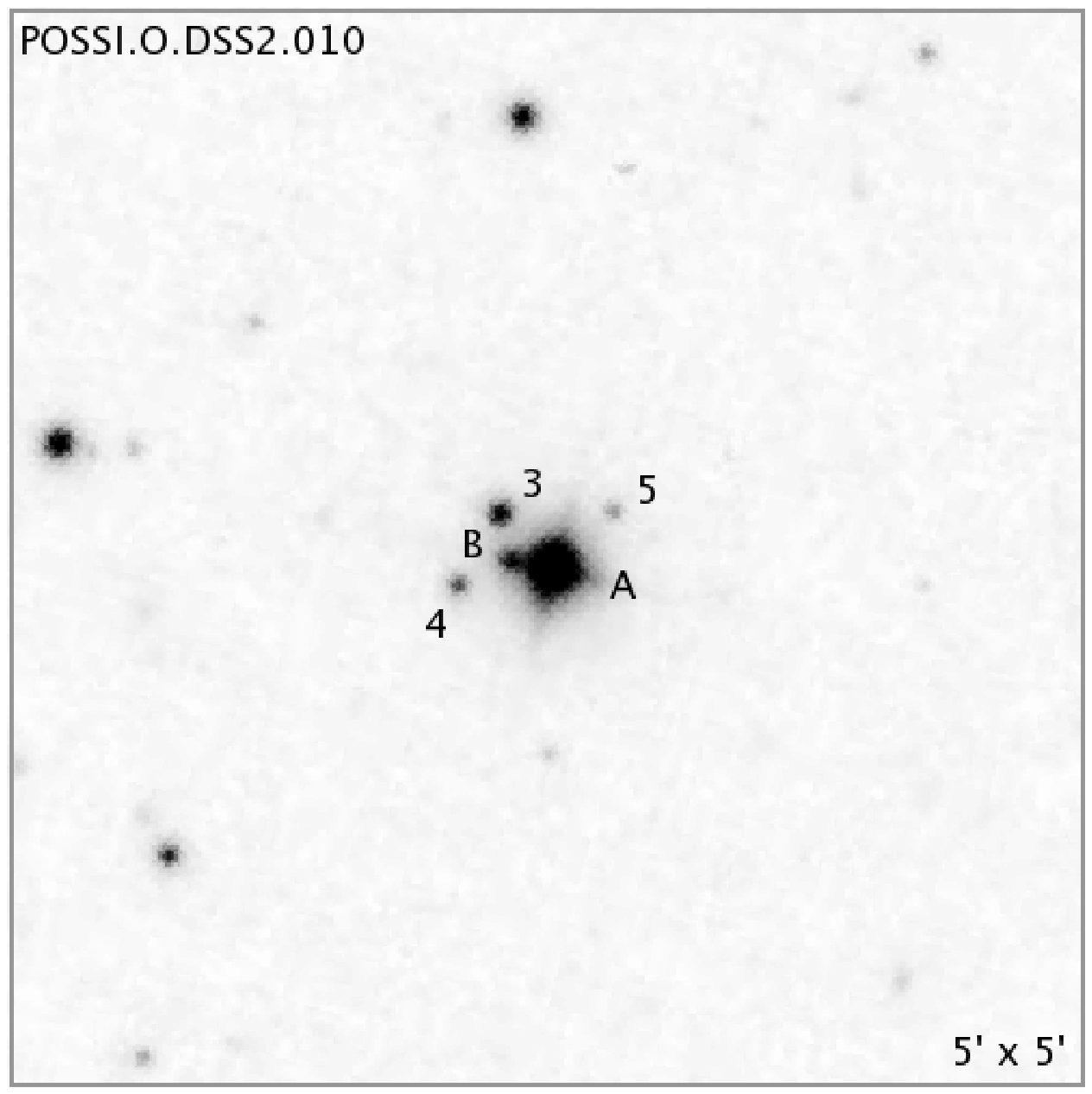,width=60mm,angle=0,clip=}}
\vspace{1mm}
\captionr{1}{Identification chart from the first Palomar Observatory
Sky Survey (POSS I.O).}
}
\hspace{.2mm}
\vtop{\hsize61mm
\centerline{\psfig{figure=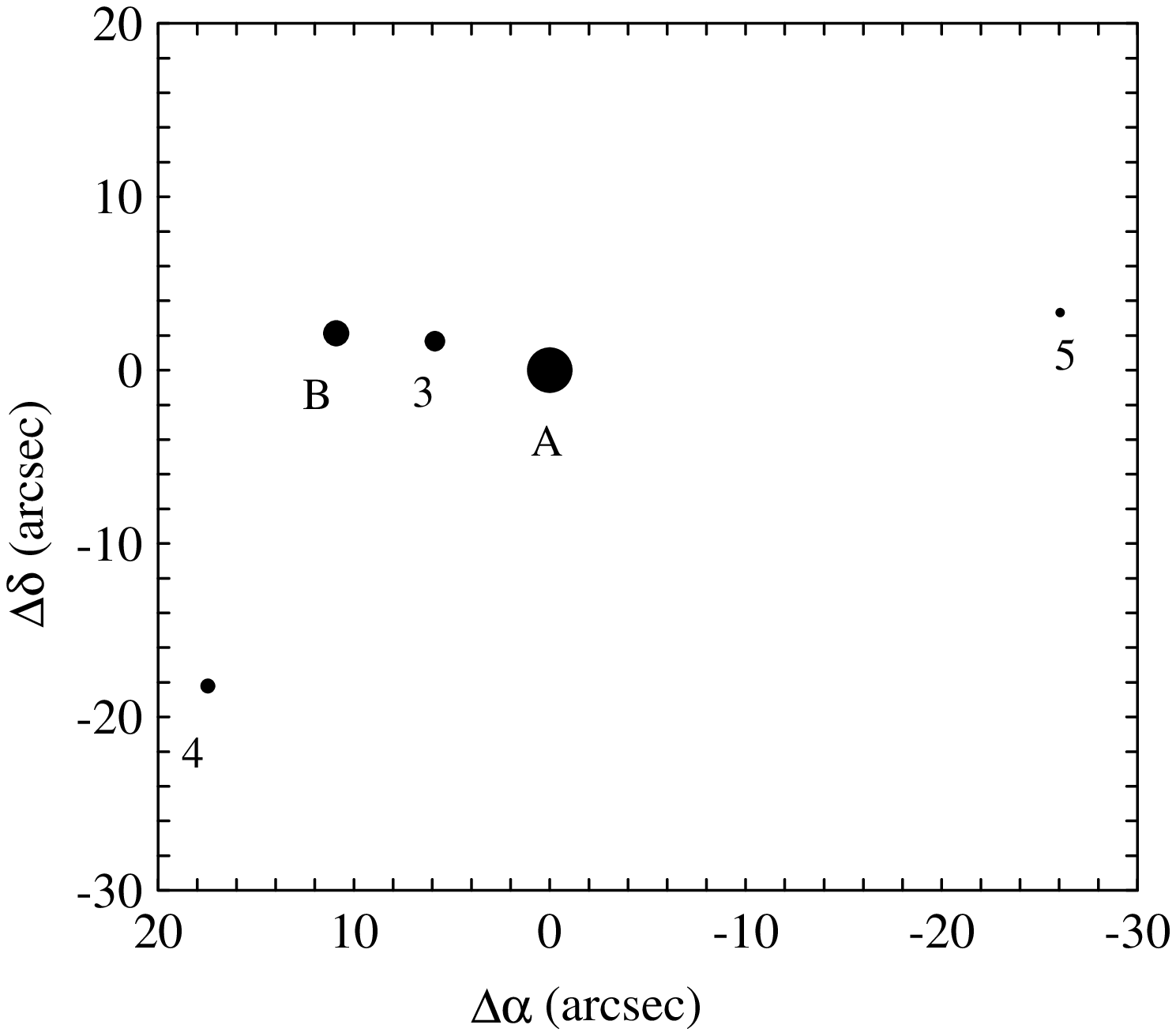,width=60mm,angle=0,clip=}}
\vspace{1mm}
\captionr{2}{Identification chart for epoch 2000.0. Proper motion
for star
3 is unknown, its position is for the epoch 1953.7.}
}}

\vskip4mm

%\begin{center}
\vbox{\footnotesize
\begin{tabular}{l}
\multicolumn{1}{c}{\parbox{124mm}{\baselineskip=8pt {\normbf Table
1.}{\norm\ ~Identification of binary
components and nearby stars. }}}\\
\tablerule
\noalign{\vskip2mm}
\noalign{Component A}
\noalign{\vskip1mm}
\noalign{BD +82 565, AGK3 +82 545, HIP 92162, PLX 4424, CCDM
J18471+8244,
WDS J18470+8244,} \noalign{IDS 18587+8236, PPM 3296, SAO 3127, G 259-37,
LTT
15562, LDS 1894, TYC 4648 474 1,} \noalign{USNO-A2.0 1725-00600458,
GSC2.2 N01001337626, 2MASS 18470262+8243300}
\noalign{\vskip2mm}
\noalign{Component B}
\noalign{\vskip1mm}
\noalign{LP 10-124, LP 9-323, 2MASS 18470873+8243320}
\noalign{\vskip2mm}
\noalign{Star 3}
\noalign{\vskip1mm}
\noalign{USNO-A2.0 1725-00600531}
\noalign{\vskip2mm}
\noalign{Star 4}
\noalign{\vskip1mm}
\noalign{USNO-A2.0 1725-00600574, GSC2.2 N01001337627, APM-N
EO0802-0246683,}
\noalign{2MASS 18471195+8243116}
\noalign{\vskip2mm}
\noalign{Star 5}
\noalign{\vskip1mm}
\noalign{USNO-A2.0 1725-00600367, APM-N EO0802-0246181, GSC2.2
N01001337628}
\tablerule
\end{tabular}
}
%\end{center}

\sectionb{3}{PHOTOMETRY AND SPECTRAL TYPES}
\vskip-11pt
From the Kharchenko (2001) ASCC-2.5 catalog the
magnitude and color index of the A component are:  $V$ =
9.32$\pm$0.02 and $B$--$V$ = 0.67$\pm$0.04.  The B component and
other three field stars have only photographic magnitudes in the
systems similar to $B$ and $R$, and they are presented in Table 2.
The infrared 2MASS photometry results available for three stars
are also given.

\begin{center}
\vbox{\footnotesize
\begin{tabular}{crcrclcl}
\multicolumn{8}{c}{\parbox{95mm}{\baselineskip=8pt
{\normbf\ \ Table 2.}{\norm\ Photometry of stars in the vicinity of BD
+82 565. }}}\\
\tablerule
Comp. & $B$~ & $\sigma_B$ & $R$~ & $\sigma_R$ & $B$--$R$ &
$\sigma_{(B-R)}$
& Source \\
\tablerule
\noalign{\vskip1mm}
   A  &  9.9\rlap{9}  &  0.03& 8.8\rlap{1} & 0.41 & 1.18 & 0.41  &  ASCC-2.5, GSC2.2 \\
   B  &  16.0  &      &14.5  &      & 1.5  &       &  NLTT  \\
   3  &  16.0  &      &14.7 &      & 1.3  &       &  USNO-A2.0 \\
   4  &  17.1\rlap{2} & 0.41 &15.2\rlap{9} & 0.42 & 1.91 & 0.59  &  GSC2.2  \\
   5  &  17.6\rlap{0} & 0.41 &17.1\rlap{5} & 0.42 & 0.45 & 0.59  &  GSC2.2  \\
\tablerule
\end{tabular}
}
\end{center}

\begin{center}
\vbox{\footnotesize
\tabcolsep = 4pt
\begin{tabular}{crccccccccc}
\multicolumn{11}{c}{\parbox{120mm}{\baselineskip=8pt {}
}}\\
\tablerule
 Comp. & $J$ & $\sigma_J$ & $H$ & $\sigma_H$ & $K$  & $\sigma_K$ &
$J$--$H$ & $\sigma_{(J-H)}$ & $H$--$K$  & $\sigma_{(H-K)}$  \\
\tablerule
\noalign{\vskip1mm}
A  & 8.085 & 0.026 & 7.773 &0.023 &  7.694 & 0.018 & 0.312 & 0.035 &  0.079 & 0.029 \\
B  &11.455 & 0.022 &10.949 &0.019 & 10.761 & 0.023 & 0.506 & 0.029 &  0.188 & 0.030 \\
4  &14.092 & 0.027 &13.487 &0.025 & 13.451 & 0.050 & 0.605 & 0.037 &  0.036 & 0.056  \\
\tablerule
\end{tabular}
}
\end{center}

Three discrepant one-dimension spectral types for the A component are
known.  Petersson (1927) in 1924 classified the star as F8 using 264
\AA/mm dispersion objective prism spectra obtained at the Uppsala
Observatory.  This type is given in the Skiff (2003) {\it Catalogue of
Stellar Spectral Classifications}.  G. P. Kuiper obtained a considerably
later spectral type, K0 (published by Bidelman (1985) in the article
``G.  P. Kuiper's spectral classifications of proper-motion stars").  A
decade earlier Bidelman \& Lee (1975) presented  Kuiper's spectral
type in a compilation of spectral types for proper motion stars pointing
Jenkins (1952) {\it Catalogue of Trigonometric Parallaxes} as the
literature
source.  In the van Altena et al.  (1995) catalog of trigonometric
parallaxes a dK0 spectral class is given, quoting Bidelman as the
literature source, so the origin of the dwarf classification is not
clear.  The third spectral class, G0, was estimated by Balz (1958) from
the McCormick Observatory 300 \AA/mm spectra, and quoted also in the
AGK3 catalog and in many other sources including the SIMBAD database.
Intrinsic color index $(B-V)_0$ = 0.65 corresponds to the G2/G5\,V
spectral type.  For the B component Luyten presented different color
classes:  from k, k-m to m. 2MASS colors correspond to a dwarf of K5--K7
spectral type.

\sectionb{4}{ DISTANCE, ABSOLUTE MAGNITUDE AND KINEMATICS }
\vskip-11pt {\it Hipparcos} recorded a good precision (6\%)
parallax $\pi$ = 13.78$\pm$0.81 mas.  This corresponds to a
distance $d$ of 72.6$\pm$4.3 pc.  Only one useless ground-based
parallax, $\pi$ = 0.0017$\pm$0.0133 mas, measured with the
Greenwich Observatory Thompson 66 cm refractor (Dyson 1925) is
included in Jenkins (1952) and van Altena et al.  (1995) Yale
General Parallax Catalogues.  The {\it Tycho} program obtained a
very negative parallax, $\pi$ = --9.90$\pm$11.20 mas.  Kharchenko
(2001), following a questionable method to average {\it Hipparcos}
and {\it Tycho} parallax determinations, has presented $\pi$ =
13.65$\pm$0.81 mas.  From Schlegel et al.  (1998) interstellar
reddening maps for the binary at $\ell$ = 114.8$\degr$ and $b$ =
27.0$\degr$ the total line-of-sight interstellar reddening is
$E_{B-V}$ = 0.07.  Taking into account a distance to the binary of
72.6 pc (from the {\it Hipparcos} parallax), the true $E_{B-V}$ =
0.02 and $A_V$ = 0.07 are calculated (Anthony-Twarog \& Twarog
1994).

The absolute magnitude of component A from the {\it Hipparcos} parallax
and the above-mentioned $V$ and $A_V$ is $M_V$ = 4.95$\pm$0.13 mag.  In
the $M_V$ vs.  $(B-V)_0$ plot the A component is situated within the
main-sequence band.  The reduced proper-motion diagram $H_V$, $B$--$V$
places the star at the subdwarf-main sequence border.  The B component
in the blue spectral region is fainter by 6 mag, in the red -- by 5.7
mag, and this corresponds to a M3/4 dwarf.  However, the infrared 2MASS
photometry of the star is consistent with an earlier K5/7 dwarf.

The {\it Hipparcos} parallax, the {\it Tycho 2} proper motion components
and our value of spectroscopic binary barycenter radial velocity are
used to calculate kinematical parameters of the system.  The procedure
of computation is the same as in Bartkevi\v cius \& Gudas (2001, 2002).
The velocity component $U$ is directed to the Galactic center, $V$ -- to
the direction of Galactic rotation and $W$ -- to the North Galactic
Pole.  They have been corrected due to the solar motion with respect to
the Local Standard of Rest $U$= 10.0$\pm$0.4 km/s, $V$ = 5.2$\pm$0.6
km/s and $W$ = 7.2$\pm$0.4 km/s (Binney \& Merrifield 1998).  Evidently,
the binary belongs to the thick disk population.

\sectionb{5}{RADIAL VELOCITY MEASUREMENTS}
\vskip-9pt
\begin{wrapfigure}[17]{l}[0pt]{77mm}
\psfig{figure=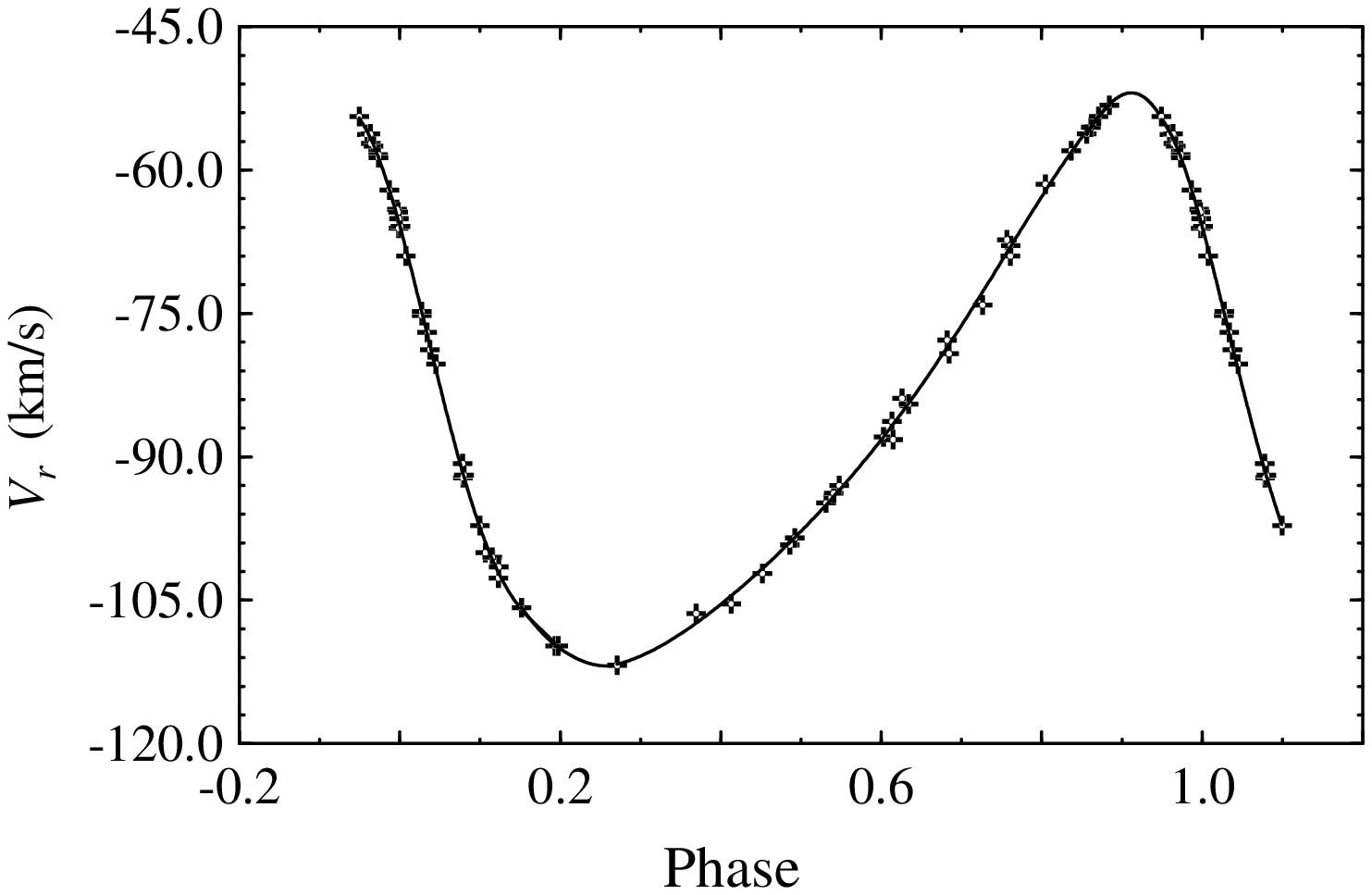,width=76mm,angle=0,clip=}
\vspace{-3mm}
\captionc{3}{Radial velocity curve.}
\end{wrapfigure}

%\newpage

Radial velocity measurements were made by J. Sperauskas with a
CORAVEL-type spectro\-meter constructed at the Vilnius University
Observatory.  A description of the measurements and data
reduction
procedures are presented in Upgren, Sperauskas \& Boyle (2002). 57
individual radial velocities for BD +82 565\,A were obtained at the
Mol\.e\-tai Observatory with the 0.63 m and 1.65 m telescopes.  These
measurements were spread over the period of 619 days starting in 2000
August 28.  Standard single-measurement errors range from 0.7 to 1.0
km/s with the mean value of 0.8 km/s.  Individual radial velocity
measurements are listed in Table 4 together with the Heliocentric Julian
Days and phases calculated from the orbital elements, measurement errors
and residuals.

\sectionb{6}{ORBITAL SOLUTION}
\vskip-9pt The obtained radial velocity curve is plotted in Figure
3. The calculated orbital elements are given in Table 5. The
system has a high center of mass radial velocity and moderate
orbit eccentricity.

\begin{center}
\vbox{\footnotesize
\begin{tabular}{crcrclclc}
\multicolumn{9}{c}{\parbox{80mm}{\baselineskip=8pt
{\normbf\ \ Table 3.}{\norm\ Kinematical parameters.}}}\\
\tablerule
\noalign{\vskip1mm}
$\ell$ & $b$ & $\mu_\alpha$ & $\sigma {\mu \alpha}$ & $\mu_\delta$ &
$\sigma {\mu \delta }$ & $V_r$ & $\sigma {V_r}$ & $V_t$ \\
degr & degr & mas & mas & mas & mas & km/s & km/s & km/s \\
114.8 & 27.0 & 183.3 & 1.5 & 292.3 & 1.5 & --86.70 & 0.07 & 118.7 \\
\tablerule
\noalign{\vskip1mm}
 $\sigma _{V_t}$ & $U$ & $\sigma _U$ & $V$ & $\sigma _V$ & $W$ &
$\sigma_W$ & $V_{\rm tot}$ & $\sigma_{V_{\rm tot}}$ \\
 km/s & km/s & km/s & km/s & km/s & km/s & km/s & km/s & km/s \\
7.0 & --62.6 & 3.2 & --84.1 & 2.8 & --84.2 & 1.2 & 147.0 & 5.6 \\
\tablerule
\end{tabular}
}
\end{center}

\begin{center}
\vbox{\footnotesize
\tabcolsep = 4pt
\begin{tabular}{ccrcr|ccrcr}
\multicolumn{10}{c}{\parbox{80mm}{\baselineskip=8pt
{\normbf\ \ Table 4.}{\norm\ Radial velocity measurements.}}}\\
\tablerule
HJD & Phase & $V_r$~~ & $\sigma V_r$ & $O-C$ & HJD & Phase & $V_r$~~ & $\sigma V_r$ &$O-C$ \\
   &        &  km/s &  km/s & km/s &     &       &  km/s & km/s & km/s \\
\tablerule
\noalign{\vskip1mm}
51785.320  &  0.486205 &     --99.2  & 0.7  &    --0.195 &  ~~52145.555  &  0.870708 &     --54.4  & 0.7  &     --0.047     \\
51794.348  &  0.197561 &    --109.8  & 0.8  &     0.377 &   ~~52147.289  &  0.007337 &     --69.0  & 0.8  &     --0.852     \\
51999.586  &  0.369167 &    --106.4  & 0.8  &    1.009 &    ~~52147.539  &  0.027036 &     --75.2  & 0.8  &     --0.402     \\
52003.557  &  0.682060 &     --77.8  & 0.9  &     0.791 &   ~~52179.320  &  0.531201 &     --94.8  & 0.7  &      0.198     \\
52004.561  &  0.761169 &     --67.9  & 0.7  &     0.268 &   ~~52180.361  &  0.613226 &     --86.3  & 0.8  &      0.360     \\
52007.564  &  0.997789 &     --64.4  & 0.8  &     0.755 &   ~~52181.266  &  0.684535 &     --79.2  & 0.8  &     --0.918     \\
52007.569  &  0.998183 &     --66.1  & 0.8  &    --0.826 &  ~~52182.234  &  0.760808 &     --69.0  & 0.8  &     --0.782     \\
52007.576  &  0.998735 &     --65.1  & 0.9  &     0.342 &   ~~52199.226  &  0.099682 &     --97.2  & 0.8  &     --0.031     \\
52007.605  &  0.001020 &     --65.9  & 1.0  &     0.244 &   ~~52203.200  &  0.412811 &    --105.4  & 0.7  &     --0.748     \\
52008.583  &  0.078081 &     --90.7  & 0.9  &     0.818 &   ~~52204.209  &  0.492315 &     --98.5  & 0.7  &     --0.016     \\
52008.589  &  0.078553 &     --92.2  & 1.1  &    --0.546 &  ~~52205.610  &  0.602706 &     --87.9  & 0.8  &     --0.094     \\
52008.603  &  0.079656 &     --92.0  & 0.9  &    --0.033 &  ~~52207.180  &  0.726413 &     --74.1  & 1.5  &    -1.225     \\
52008.609  &  0.080129 &     --91.9  & 0.9  &     0.201 &   ~~52207.563  &  0.756591 &     --67.3  & 1.4  &     1.496     \\
52032.330  &  0.949211 &     --54.4  & 0.7  &     0.070 &   ~~52208.578  &  0.836568 &     --58.0  & 0.9  &      0.040     \\
52032.455  &  0.959061 &     --56.3  & 0.7  &    --0.386 &  ~~52210.188  &  0.963427 &     --56.2  & 0.8  &      0.477     \\
52032.556  &  0.967019 &     --57.5  & 0.8  &    --0.138 &  ~~52210.195  &  0.963978 &     --57.2  & 0.8  &     --0.421     \\
52033.321  &  0.027297 &     --74.8  & 0.7  &     0.088 &   ~~52210.310  &  0.973040 &     --58.4  & 0.8  &      0.222     \\
52033.405  &  0.033915 &     --77.0  & 0.7  &     0.186 &   ~~52210.315  &  0.973434 &     --58.7  & 0.8  &      0.010     \\
52033.449  &  0.037382 &     --78.8  & 0.8  &    --0.412 &  ~~52210.484  &  0.986750 &     --62.1  & 0.8  &     --0.107     \\
52033.545  &  0.044947 &     --80.3  & 0.7  &     0.692 &   ~~52375.594  &  0.996494 &     --64.1  & 0.8  &      0.666     \\
52034.329  &  0.106721 &    --100.0  & 0.7  &   -1.230 &    ~~52377.566  &  0.151877 &    --105.8  & 0.9  &      0.561     \\
52034.437  &  0.115231 &    --100.5  & 0.8  &     0.048 &   ~~52382.585  &  0.547346 &     --93.0  & 0.8  &      0.463     \\
52034.534  &  0.122874 &    --102.7  & 0.7  &    --0.699 &  ~~52383.580  &  0.625746 &     --83.9  & 0.8  &     1.367     \\
52034.539  &  0.123268 &    --101.5  & 0.8  &     0.572 &   ~~52386.568  &  0.861184 &     --55.5  & 0.8  &     --0.240     \\
52089.473  &  0.451760 &    --102.2  & 0.8  &    --0.401 &  ~~52398.540  &  0.804511 &     --61.5  & 0.7  &      0.713     \\
52141.368  &  0.540796 &     --93.8  & 0.8  &     0.292 &   ~~52399.554  &  0.884408 &     --53.2  & 0.7  &      0.082     \\
52142.305  &  0.614626 &     --88.2  & 0.8  &   -1.694 &    ~~52403.478  &  0.193598 &    --109.8  & 0.7  &      0.166     \\
52142.549  &  0.633852 &     --84.5  & 0.7  &    --0.153 &  ~~52404.457  &  0.270737 &    --111.8  & 0.7  &     --0.588     \\
52145.367  &  0.855895 &     --56.2  & 0.7  &    --0.390 &             &          &            &          &     \\
\tablerule
\end{tabular}
}
\end{center}

\begin{center}
\vbox{\footnotesize
\begin{tabular}{ll}
\multicolumn{2}{c}{\parbox{80mm}{\baselineskip=8pt
{\normbf\ \ Table 5.}{\norm\ Orbital elements of  BD +82 565A.}}}\\
\tablerule
Parameter & Value  \\
\tablerule
\noalign{\vskip1mm}
        Orbital period  &          $P$  =  12.6913$\pm$0.0009 days        \\
        Center-of-mass velocity & $V_0$ =  --86.70$\pm$0.07 km/s     \\
        Half-amplitude          &  $K$ =   29.55$\pm$0.10 km/s       \\
        Eccentricity           &   $e$ = 0.305$\pm$0.003             \\
        Longitude of periastron & $\omega$ = (57.3$\pm$0.6)$\degr$ \\
  Date of conjunction & $T_{\rm conj}$ = 2452401.021$\pm$0.022 HJD   \\
Projected semimajor axis &  $a \sin i$ = (4.91$\pm$0.02)\,10$^6$ km     \\
        Function of the mass &  $f(m)$ = 0.0294$\pm$0.0004 $M_{\odot}$ \\
        Mean square error of one observation & $\sigma_(O-C) = \pm$ 0.44 km/s \\
\tablerule
\end{tabular}
}
\end{center}

\sectionb{7}{VISUAL SUBSYSTEM PARAMETERS}
\vskip-9pt
The period of the AB subsystem of almost 17\,000 years
is estimated using Kepler's third law, assuming circular face-on
orbit, apparent separation $d$ = 11.5\arcsec, parallax $\pi$ =
13.78 mas and total mass 1.8 $M_{\odot}$, adopting for a
spectroscopic binary A the main component $M$ = 1 $M_{\odot}$
(according to its spectral class) and for the secondary component
0.5 $M_{\odot}$ (from the spectroscopic mass function $f(m)$
taking $\sin^3 i$ = 2/3). For the visual B component we adopted
0.3 $M_{\odot}$
 (from the mass-luminosity relation).  From the
Palomar first and second epoch Sky Surveys and 2MASS survey crude
estimates of the angular separation and position angle were made for two
epochs.  For $E_{\rm mean}$ = 1953.7 we obtain:  $d_{\rm mean}$ =
(11.18$\pm$0.21)\arcsec\ and {\it PA}$_{\rm mean}$ =
(79.65$\pm$0.69)$\degr$
and for $E_{\rm mean}$ = 1998.1 we obtain $d_{\rm mean}$ =
(11.46$\pm$0.07)\arcsec\ and {\it PA}$_{\rm mean}$ =
(83.07$\pm$0.33)$\degr$.
Evidently, during 44 years the angular separation practically did not
change.  Only the change of the position angle of 0.077$\degr$ per year
may be real.  In case of the circular orbit, this change of {\it PA}
corresponds to a period of about 4700 years which is almost four times
smaller than that calculated from the third Kepler law.

The minimum spatial distance between components A and B, adopting the
projected spatial distance from the mean angular separation $d$ =
11.4\arcsec\ and a distance of 72.6 pc, is $\sim$\ 830 AU.

\sectionb{8}{CONCLUSIONS}
\vskip-9pt
A short-period ($P$ = 12.69~d) and moderate
eccentricity ($e$ = 0.30) spectroscopic orbit of the A component
of a high velocity ($v_{\rm tot}$ = 147.0 km/s) visual binary
system BD +82 565 is determined from 57 CORAVEL-type radial
velocity measurements.  The projected spatial separation of
components of the visual binary AB is $\sim$\ 830 AU.

\vskip5pt

ACKNOWLEDGMENTS. We are indebted to V.-D.  Bartkevi\v cien\. e for
preparation of the manuscript to publication.  In this
investigation the information from the Strasbourg Stellar Data
Center (CDS), NASA Bibliographic Data Center (ADS), Astrophysics
preprint archive and the Washington Visual Double Stars Catalog
(WDS) were used. Radial velocity observations were obtained with
the 0.63 m and 1.65 m telescopes of the Mol\.etai Observatory,
Lithuania.

\References

\refb Anthony-Twarog B. J., Twarog B. A. 1994, AJ, 107, 1577

\refb Balz A.\,G.\,A. 1958, Publ. McCormick Obs., 13, 1

\refb Bartkevi\v cius A., Gudas A. 2001, Baltic Astronomy, 10, 481

\refb Bartkevi\v cius A., Gudas A. 2002, Baltic Astronomy, 11, 153

\refb Bartkevi\v cius A., Sperauskas J. 1990, in {\it Proceedings of the
Nordic -- Baltic Astro\-nomy Meeting}, eds.  C.-I.  Lagerkvist, D.
Kiselman \& M. Lindgren, Astronomical Observatory of the Uppsala
University, p. 45

\refb Bartkevi\v cius A., Sperauskas J. 1994, Baltic Astronomy, 3, 49

\refb Bartkevi\v cius A., Sperauskas J. 1999, Baltic Astronomy, 8, 325

\refb Bartkevi\v cius A., Sperauskas J. 2005, Baltic Astronomy, 14, 511
(this issue)

\refb Bartkevi\v cius A., Sperauskas J., Rastorguev A. S., Tokovinin
A. A. 1992, Baltic Astronomy, 1, 47

\refb Bidelman W. P. 1985.  {\it G. P. Kuiper's Spectral Classifications
of Proper-motion Stars},  ApJS, 59, 197

\refb Bidelman W. P., Lee S.-G. 1975, AJ, 80, 239

\refb Binney J., Merrifield M. 1998  {\it Galactic Astronomy},
Princeton University Press, Princeton

\refb Carrington R, C. 1857, {\it A Catalogue of 3735 Circumpolar Stars,
observed at Redhill in the Years 1854, 1855, and 1856, and reduced to
Mean Positions for 1855.0}, London

\refb Dyson F. 1925, {\it Greenwich Observations of Stellar Parallaxes}

\refb Giclas H. L., Burnham R., Thomas N. G. 1970, Bull.  Lowell Obs., 8
(No. 152), 165

\refb Jenkins L. F. 1952, {\it General Catalogue of Trigonometric
Stellar Parallaxes},  Yale University Observatory,  New Haven

\refb Kharchenko N. V.
2001,  {\it All-sky Compiled Catalogue of 2.5 Million Stars},
Kinematics and Physics of Celestial Bodies, Kiev, 17, 409;
CDS Catalog No.I/280A

\refb Luyten W. J. 1961. {\it A Catalogue of 7127 Stars in the Northern
Hemisphere with Proper Motions Exceeding 0.2\arcsec\ Annually},
Minneapolis

\refb Luyten W. J. 1966,  Publ.  Minnesota Obs., No. 17

\refb Luyten W. J. 1967, {\it Proper Motion Survey with the 48-inch
Schmidt Telescope.  XII}.  Minnesota University Observatory, Minneapolis

\refb Luyten W. J. 1968, {\it Proper Motion Survey with the 48-inch
Schmidt Telescope.  XIV}, Minnesota University Observatory, Minneapolis

\refb Luyten W. J. 1969. {\it Proper Motion Survey with the 48-inch
 Schmidt Telescope.  XXI},  Minnesota University Observatory,
Minneapolis;  CDS Catalog No.  I/130

\refb Luyten W. J. 1979.  {\it NLTT Catalogue.  Vol. I. +90 to +30},
University of Minnesota,  Minneapolis;  CDS Catalog No. I/98A

\refb Udry S., Mayor M., Maurice E., Andersen J., Imbert M., Lindgren
H., Mermilliod J. C. 1999, in {\it Precise Stellar Radial Velocities},
eds.  J. B. Hearnshaw \& C. D. Scarfe, ASP Conf.  Ser., 185, 383 (IAU
Colloq.  No. 170)

\refb Upgren A. R., Sperauskas J., Boyle R. P. 2002, Baltic Astronomy,
11, 91

\refb Petersson J. H. 1927, Medd.  Uppsala Obs., 29, 1

\refb Salim S., Gould A. 2003, ApJ, 582, 1011; CDS Catalog No.
J/ApJ/582/1011

\refb Schlegel D. J., Finkbeiner D. P., Davis M. 1998, ApJ, 500, 525

\refb Skiff B. A. 2003.  {\it Catalogue of Stellar Spectral
Classifications}, Lowell Observatory,  CDS Catalog No.  III/233

\refb Sperauskas J., Bartkevi\v cius A. 2002, AN, 323, 139

\refb van Altena W. F., Lee J. T., Hoffleit E. D. 1995, {\it The General
Catalogue of Trigonometric Stellar Parallaxes}, 4th edition, Yale
University Observatory

 \end{document}